\documentclass[prc,reprint,twocolumn,amsmath,amssymb,superscriptaddress]{revtex4-1}
\usepackage{graphicx}
\usepackage{color}

\newcommand{\be}{\begin{equation}}
\newcommand{\ee}{\end{equation}}
\newcommand{\beq}{\begin{eqnarray}}
\newcommand{\eeq}{\end{eqnarray}}
\newcommand{\ba}{\begin{array}}
\newcommand{\ea}{\end{array}}

\begin{document}

\title{Is the LHCb $P_c(4312)^+$ plausible in the GlueX $\gamma p\to J/\psi p$ total cross sections ?}

\vspace{3mm}
\date{\today}

\author{\mbox{Igor~Strakovsky}}
\altaffiliation{Corresponding author: \texttt{igor@gwu.edu}}
\affiliation{Institute for Nuclear Studies, Department of Physics, 
The George Washington University, Washington, DC 20052, USA}

\author{\mbox{William~J.~Briscoe}}
\affiliation{Institute for Nuclear Studies, Department of Physics, The
George Washington University, Washington, DC 20052, USA}

\author{\mbox{Eugene~Chudakov}}
\affiliation{Thomas Jefferson National Accelerator Facility, Newport News, Virginia 23606, USA}

\author{\mbox{Ilya~Larin}}
\affiliation{University of Massachusetts (UMASS Amherst), Amherst, MA 01003, USA}
\affiliation{National Research Centre ``Kurchatov Institute,"  Moscow 117218, Russia}

\author{\mbox{Lubomir~Pentchev}}
\affiliation{Thomas Jefferson National Accelerator Facility, Newport News, Virginia 23606, USA}

\author{\mbox{Axel~Schmidt}}
\affiliation{Institute for Nuclear Studies, Department of Physics, 
The George Washington University, Washington, DC 20052, USA}

\author{\mbox{Ronald~L.~Workman}}
\affiliation{Institute for Nuclear Studies, Department of Physics, 
The George Washington University, Washington, DC 20052, USA}

\noaffiliation

\begin{abstract}
New high-statistics total cross section data for $\gamma p\to J/\psi p$ from the GLUonic EXcitation (GlueX) experiment are fitted in a search for the exotic  $P_c(4312)^+$ state observed by the Large Hadron Collider beauty (LHCb) collaboration. The integrated luminosity of this GlueX experiment was about $320~\mathrm{pb^{-1}}$. The fits show that destructive interference involving an $S$-wave resonance and associated non-resonance background produces a sharp dip structure about $75~\mathrm{MeV}$ below the LHCb mass, in the same location as a similar structure is seen in the data. Limitations of the employed model and the need for improved statistics are discussed.
\end{abstract}

\maketitle

\clearpage

The formation of hadrons from quarks and gluons is governed by quantum chromodynamics (QCD). 
The traditional quark model~\cite{Gell-Mann:1964ewy, Zweig:1964jf}, where 
hadrons are classified as mesons (composed of $q\overline{q}$) and baryons (composed of $qqq$), has provided a satisfactory description of the hadrons observed over the last 75 years. However, the last two decades have witnessed the emergence of states or resonant structures in experiments, many of which do not fit the hadron spectrum predicted by the naive quark model and thus are candidates to be exotic states. In particular, the  Large Hadron Collider beauty (LHCb) experiment has reported the observation of four narrow $P_c^+$ pentaquark states in decays 
    $$\Lambda_b^0\to J/\psi + p + K^- \>,$$ 
    $$\Lambda_b^0\to J/\psi + p + \pi^- \>,$$ 
and 
$$B_s^0\to J/\psi + p + \overline{p}$$ 
with hidden heavy charm quark contributions~\cite{Aaij:2015tga, LHCb:2016lve, Aaij:2019vzc, LHCb:2021chn}.  The masses of these four narrow $P_c$ states, as well as their significance, are given in Table~\ref{tab:tbl1}. In response, many theoretical analyses evaluated these cases (see for instance, Refs.~\cite{Wang:2015jsa, Kubarovsky:2015aaa, Karliner:2015voa, Blin:2016dlf, Eides:2019tgv, Semenova:2019gzf, Burns:2021jlu, Peng:2019wys, Zhang:2023czx}). In particular, in
Refs.~\cite{Wang:2015jsa, Kubarovsky:2015aaa, Karliner:2015voa, Blin:2016dlf}, it was proposed to search for this $P_c$ state in the $J/\psi$ photoproduction off the nucleon. The corresponding cross section was evaluated based on the Vector Meson Dominance (VMD) model~\cite{Gell-Mann:1961jim, Kroll:1967it, Sakurai:1969}.

\begin{table}[htb!]
\caption{Summary of $P_c^+$ properties reported by the LHCb
    Collaboration~\protect\cite{Aaij:2019vzc, LHCb:2021chn}. \label{tab:tbl1}}
\begin{center}
\begin{tabular}{|c|c|c|c|}
\hline
   State &   M   & $\Gamma[Pc\to J/\psi + p]$ & Significance \\
         & (MeV) & (MeV)                      & \\
\hline
$P_c(4312)^+$ & $4311.9\pm 0.7_{-0.6}^{+6.8}$  & $9.8\pm 2.7_{-4.5}^{+3.7}$       & 7.3~$\sigma$ \\
$P_c(4337)^+$ & $4337{^{+7}_{-4}}{^{+2}_{-2}}$ & $29{^{+26}_{-12}}{^{+13}_{-14}}$ & 3.1 -- 3.7~$\sigma$ \\
$P_c(4440)^+$ & $4440.3\pm 1.3_{-4.7}^{+4.1}$  & $20.6\pm 4.9_{-10.1}^{+8.7}$     & 5.4~$\sigma$ \\
$P_c(4457)^+$ & $4457.3\pm 0.6_{-1.7}^{+4.1}$  & $6.4\pm 2.0_{-1.9}^{+5.7}$       & 5.4~$\sigma$ \\
\hline
\end{tabular}
\end{center}
\end{table}

\begin{figure}[ht]
\centering
{
    \includegraphics[width=0.25\textwidth,keepaspectratio]{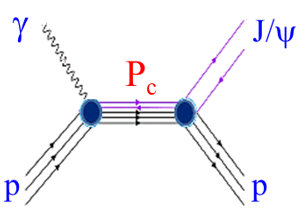}
}

\centerline{\parbox{0.70\textwidth}{
    \caption[] {\protect\small 
    $s$-channel resonance $P_c$.} 
    \label{fig:fig1} } }
\end{figure}

The GLUonic EXcitation (GlueX) experiment measured total cross sections for the reaction $\gamma + p\to J/\psi + p$ and reported no evidence for the LHCb pentaquark candidates $P_c^+$ (Fig.~\ref{fig:fig1}) and set model-dependent upper limits on their branching ratios (BRs)~\cite{Ali:2019lzf}.

These interpretations of the LHCb and GlueX cases considered a constructive interference between a resonance and non-resonant background for particular wave, $L^J$ (here, $L$ is a total orbital momentum and $J$ is the total angular momentum). In this work, we relax this requirement and consider the possibility of destructive interference as well.

\begin{figure}[ht]
\vspace{0.4cm}
\centering
{
    \includegraphics[width=0.4\textwidth,angle=90,keepaspectratio]{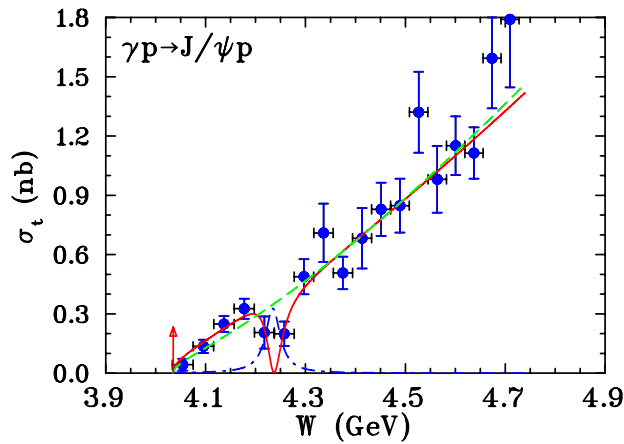}
}

\centerline{\parbox{0.50\textwidth}{
    \caption[] {\protect\small 
    Best-fit results for new GlueX total cross sections for the reaction $\gamma + p\to J/\psi + p$ (blue filled circles)~\cite{GlueX:2023}. The vertical error bars represent the total uncertainties (statistical and point-to-point systematic uncertainties in quadrature). The horizontal error bars reflect the energy binning (not used in the fit). The best-fit result, using Eq.~(\protect\ref{eq:eq3}), shown by red solid curve.  Green dashed curve corresponds to the non-resonant fit as a function of $q$ (Eq.~(\protect\ref{eq:eq4a})). Blue dash-dotted curve corresponds to the $S$-wave resonance. The red vertical arrow indicates the $J/\psi$ production threshold ($W = 4.035~\mathrm{GeV}$).
    } \label{fig:fit5} } }
\end{figure}

New higher-statistics GlueX measurements of the threshold total cross section, for the reaction $\gamma + p\to J/\psi + p$~\cite{GlueX:2023}, motivate an alternative search for the LHCb exotics. The integrated luminosity of this recent GlueX experiment was about $320~\mathrm{pb^{-1}}$, a large improvement over the approximately $68~\mathrm{pb^{-1}}$ of the GlueX data presented in Ref.~\cite{Ali:2019lzf}). Within the limits of the available statistics, the clearest structure is a dip observed in the recent GlueX total $\gamma + p\to J/\psi + p$ cross sections, in vicinity of $W = 4.3~\mathrm{GeV}$~\cite{GlueX:2023} (Fig.~\ref{fig:fit5}), a possible indication of the lowest-lying LHCb $P_c(4312)$~\cite{Aaij:2019vzc}. This dip becomes more pronounced in new GlueX data~\cite{GlueX:2023} versus the initial GlueX report~\cite{Ali:2019lzf}.
We therefore consider whether simple background/resonance interference is capable of reproducing the low-energy behavior visible in the cross section data. Through interference, the cross section can develop either a peak or a dip. Pedagogical reviews of quantum interference are given in Refs.~\cite{Azimov:2009ta, LeviSetti:1973}.

A general expressions for the total cross section of the inelastic binary reaction, $\gamma + p\to J/\psi + p$, can be written as
\begin{equation}
	\sigma_t = \int_0^{2\pi}\int_0^\pi\frac{d\sigma}{d\Omega}~\sin\theta~ d\theta~d\phi \>,
    \label{eq:eq1}
\end{equation}
where $\theta$ and $\phi$ are the $J/\psi$ polar and azimuthal production angles, respectively.

Phenomenologically, the total cross section ($\sigma_t$), using the Landau-Lifshitz normalisation,~\cite{Landau:1991wop} is
\begin{equation}
	\sigma_t = \frac{\pi}{4k^2} \sum_{J=0}^{\infty}(2J + 1)~|f|^2 \>,
    \label{eq:eq2}
\end{equation}
where $k$ is the photon momentum in the $\gamma p$ center-of-mass (c.m.) system.

For an $s$-channel resonance in the reaction $\gamma + p\to P_c^+\to J/\psi + p$, in the considered partial wave, $L^J$,(Fig.~\ref{fig:fig1}), we shall separate the non-resonant ($b$) and resonance ($R$) parts in the partial amplitude ($f$) following Ref.~\cite{Strakovsky:1984hb, LeviSetti:1973}. Then for the dominant partial wave, we can write
\begin{equation}
	f = b + R\cdot \exp(2i\alpha) \>,
    \label{eq:eq3}
\end{equation}
where $\alpha$ is the relative phase shift which is responsible for the interference between resonance ($R$) and non-resonant ($b$) components of the partial amplitude. The interference may be either positive (constructive) or negative (destructive). We make $\alpha$ a free parameter in our fit of the cross sections, not given by any theory. Let us note that the resonance can interfere differently in different decay channels, at least due to different properties of the corresponding backgrounds.

For the resonance part of the partial amplitude, $f$, we shall use a canonical relativistic Breit-Wigner parameterization:
\begin{equation}
		R = \frac{2\Gamma M}{[(M)^2 - s] - i\Gamma M}~X~\>, 
    \label{eq:eq5}
\end{equation}
where $s$ is the square of the total energy in c.m. Then, $M$, $\Gamma$, and $X$ are respectively the mass and total and partial widths
\begin{equation}
	X = \frac{\sqrt{\Gamma(\gamma + p)~\Gamma(J/\psi + p)}}{\Gamma} 
	\nonumber 
\end{equation}
\begin{equation}
	~~~~= \sqrt{X(\gamma + p)~X(J/\psi + p)} \>
    \label{eq:eq6}
\end{equation}
of a particular resonance. Here, $\Gamma(\gamma + p)$ and $\Gamma(J/\psi + p)$ are the partial decay widths of the resonance $P_c$ in to two-particle channels, $P_c\to \gamma + p$ and $P_c\to J/\psi + p$, respectively. Since the width $\Gamma$ of $P_c$ is rather small, we shall assume a width with no energy dependence
avoiding the need to modify the Breit-Wigner parameterization with additional form factors, leaving just three free
resonance parameters $M$, $\Gamma$, and $X$ to fit.

Traditionally, the total cross section behavior vs energy of the near-threshold binary inelastic reaction $m_a + m_b < m_c + m_d$ is described as series of \textit{odd} powers in $q$ (\textit{even} powers in the case of elastic scattering), as we used for the $J/\psi-p$ scattering length determination recently~\cite{Strakovsky:2019bev}:
\begin{equation}
		\sigma_t = A~q + B~q^3 + C~q^5 \>,
    \label{eq:eq4}
\end{equation}
where $A$, $B$, and $C$ are free parameters and $q$ is the c.m. momentum of the vector meson in the $J/\psi-p$ system. The \textit{linear} term is determined by two independent $S$-waves only with the total spin 1/2 and/or 3/2. Contribution to the \textit{cubic} term comes from both $P$-wave amplitudes and energy dependence of $S$-wave amplitudes. Then, the \textit{fifth}-order term arises from $D$-waves and energy dependencies of $S$- and $P$-waves. 
The phenomenological study of the $J/\psi-p$ scattering length, using GlueX data~\cite{Ali:2019lzf}, has shown that parameters $B$ and $C$ parameters are consistent with zero (within uncertainties)~\cite{Strakovsky:2019bev}. For that reason, for the background (non-resonant) amplitude, we will use just a \textit{linear} and \textit{cubic} terms from Eq.~(\ref{eq:eq4}):
\begin{equation}
		b = \sqrt{A~q~+~B~q^3}  \>,
    \label{eq:eq4a}
\end{equation}


In general, three other (higher mass) $P_c$s should reveal themselves in the $\gamma+p\to J/\psi + p$ cross section.
This would require better statistics and a finer energy binning. If the GlueX Collaboration treats the two total cross section points in the vicinity of $W = 4.3~\mathrm{GeV}$ as a potential dip (taking into account statistical and point-to-point systematic uncertainties), the probability that they are not a statistical fluctuation from an underlying smooth fit to the observed cross sections corresponds to a significance of $2.6~\mathrm{\sigma}$~\cite{GlueX:2023}. However, if one considers the probability for any two adjacent points in the whole energy interval ($W = 4.05-4.71~\mathrm{GeV}$) to have a deviation of at least this size, the significance reduces to $1.4~\mathrm{\sigma}$~\cite{GlueX:2023}.  

Because the current statistics of the GlueX measurements $2270\pm 58$ data~\cite{GlueX:2023} is by a factor of about 5 higher than recently published $469\pm 22$~\cite{Ali:2019lzf}, let us look for the effect of the LHCb low-lying $P_c$ through the destructive interference of the $S$-wave resonance and associated non-resonance background.


\begin{table}[htb!]
\caption{Best-fit results for the new GlueX data using Eq.~(\ref{eq:eq2}). 2nd column presents results for statistical and point-to-point systematic uncertainties which are combined in quadrature.
\label{tab:tbl2}}
\begin{center}
\begin{tabular}{|c|c|c|}
\hline
Quantity & Units                    & Parameter \\
\hline
M        & $\mathrm{MeV}$           & 4235$\pm$ 8 \\
$\Gamma$ & $\mathrm{MeV}$           &   35.4$\pm$ 8.2 \\ 
X        &                          &    0.023$\pm$ 0.005 \\
$\alpha$ & $\mathrm{deg}$           &   40.8$\pm$ 5.7 \\
A        & $\mathrm{nb\cdot GeV/c}$ &    0.00251$\pm$ 0.00046  \\
B        & $\mathrm{nb/(GeV/c)}$    &    0.00688$\pm$ 0.00083 \\
\hline
\end{tabular}
\end{center}
\end{table}

\begin{figure}[ht]
\vspace{0.4cm}
\centering
{
    \includegraphics[width=0.4\textwidth,angle=90,keepaspectratio]{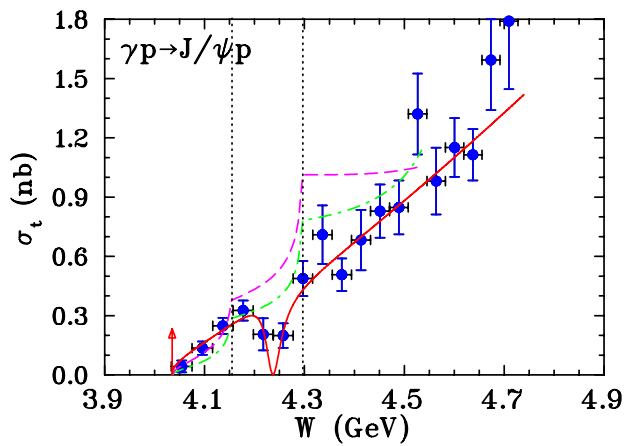}
}

\centerline{\parbox{0.50\textwidth}{
    \caption[] {\protect\small 
    The GlueX total cross section for the $\gamma + p\to J/\psi + p$ photoproduction (blue filed circles)~\protect\cite{GlueX:2023}. The open charm model predictions~\protect\cite{Du:2020bqj} shown by magenta dashed (green dash-dotted) curves with $q_{max}$ = 1~GeV/$c$ ($q_{max}$ = 1.2~GeV/$c$). 
    This model does not fit the GlueX data and has no normalization factor. 
    The phenomenological best-fit result using Eq.~(\protect\ref{eq:eq3}) shown by red solid curve.  Vertical black dotted lines show $\Lambda_c\bar{D}^{(\ast)}$-thresholds. The red vertical arrow indicates the $J/\psi$ production threshold ($W = 4.035~\mathrm{GeV}$).} 
    \label{fig:fit6} } }
\end{figure}

The best-fit results for new GlueX total cross sections for the reaction $\gamma + p\to J/\psi + p$~\cite{GlueX:2023}, using Eq.~(\ref{eq:eq2}) is given in Table~\ref{tab:tbl2} and is shown on Figs.~\ref{fig:fit5} and \ref{fig:fit6}. Additionally, we found a sensitivity to the relative phase shift, $\alpha$.

Excluding a resonance contribution from Eq.~(\ref{eq:eq3}),
one can get the following result for the non-resonant contribution, using Eq.~(\ref{eq:eq4a}): \\
$A =  (0.00183\pm  0.00040)~\mathrm{nb\cdot GeV/c}$~, \\
$B =  (0.00766\pm  0.00077)~\mathrm{nb/(GeV/c)}$~, \\
There is reasonable agreement between non-resonant parameters of two fits.

There were 6 free parameters for the $P_c(4312)^+$ resonance in the $S$-wave case and the overall $\mathrm{\chi^2/ndf}$ = 11.99/12 = 1.00. The best-fit for the alternative hypothesis (just non-resonant case) gives $\mathrm{\chi^2/ndf}$ = 19.74/16 = 1.23. The $A$ parameter is very small and in agreement with a very small $J/\psi-p$ scattering length determined recently~\cite{Strakovsky:2019bev, Pentchev:2020kao, Du:2020bqj}. Overall, our phenomenology does not allow us to select a particular partial wave for the $P_c(4312)^+$ resonance ($S$- or $P$-wave). To get a $P$-wave $P_c(4312)$ resonance partial width, one needs $X/\sqrt{3}$ and $A/\sqrt{3}$ with $B/\sqrt{3}$.


Here we have shown that a resonance-like structure is ``plausiable'' in the GlueX data~\cite{GlueX:2023},
in an energy region close to the low-mass LHCb pentaquark~\cite{Aaij:2019vzc, LHCb:2021chn}. The shift in masses between GlueX and LHCb results ($77~\mathrm{MeV}$) may depend on the reaction mechanism (including cusps (open charm) and background choices). One
should note that if a ``bump'' is imposed on the GlueX data ``by hand'' (consider the 7th - 9th energy values up from threshold),
a qualitative description of the data up to $4.35~\mathrm{GeV}$ is possible, but with a higher chi-squared, if the above fit form is used.
The statistics of the present GlueX data shown in Figs.~\ref{fig:fit5} and \ref{fig:fit6} is not sufficient to draw a definite conclusion.

We conclude with an admission that the 
analytical structure in the vicinity of the LHCb $P_c$ resonances is not as simple as a resonance plus background treatment would suggest. There are additional contributions to the $P_c$ effect associated with the open charm impact~\cite{Du:2020bqj}. In particular, the $\Lambda_c\bar{D}^{(\ast)}$ -- cusp effects could be visible. Actually, the interference between open charm and gluon exchange may (by some accident) produce a dip (see Fig.~\ref{fig:fit6}) but there is room for the resonance. The green dash-dotted curve ($q_{max}$ = 1.2~GeV/$c$) reasonably agrees with the new GlueX data. While not evident in the GlueX data, one cannot exclude that we have all 4 LHCb $P_c$ resonances~\cite{Aaij:2019vzc} together with open charm and gluon exchange (gluon contribution can be strongly suppressed due to the ``young'' effect~\cite{Feinberg:1980yu}, details about ``young'' effect in the vector meson and nucleon interaction at the threshold is given in Ref.~\cite{Strakovsky:2021vyk}). Complicated interference between these different components should motivate new high statistics experiments.


We thank Misha Ryskin, Misha Voloshin, Anatoly Dolgolenko, Moskov Amaryan, and Michael Eides for useful remarks and continuous interest in the paper and Feng-Kun Guo with Meng-Lin Du for $J/\psi p$ cross section predictions. This work was supported in part by the U.~S. Department of Energy, Office of Science, Office of Nuclear Physics, under Awards No.~DE--SC0016583 and DE--SC0016582 and Contract No.~DE-–AC05-–06OR23177.



\begin{thebibliography}{99}
\bibitem{Gell-Mann:1964ewy}
    M.~Gell-Mann,
    ``A schematic model of baryons and mesons,''
    Phys.\ Lett.\ \textbf{8}, 214 (1964).
\bibitem{Zweig:1964jf}
    G.~Zweig,
    ``An SU(3) model for strong interaction symmetry and its breaking,''
    Preprint CERN--TH--412 (1964).
\bibitem{Aaij:2015tga}
  R.~Aaij \textit{et al.} [LHCb Collaboration],
  ``Observation of $J/\psi p$ resonances consistent with pentaquark states in $\Lambda_b^0 \to J/\psi K^- p$ decays,''
  Phys.\ Rev.\ Lett.\ \textbf{115}, 072001 (2015).
\bibitem{LHCb:2016lve}
  R.~Aaij \textit{et al.} [LHCb Collaboration],
  ``Evidence for exotic hadron contributions to $\Lambda_b^0 \to J/\psi p \pi^-$ decays,''
  Phys.\ Rev.\ Lett.\ \textbf{117}, 082003 (2016).
\bibitem{Aaij:2019vzc}
  R.~Aaij \textit{et al.} [LHCb Collaboration],
  ``Observation of a narrow pentaquark state, $P_c(4312)^+$, and of two-peak structure of the $P_c(4450)^+$,''
  Phys.\ Rev.\ Lett.\ \textbf{122}, 222001 (2019).
\bibitem{LHCb:2021chn}
    R.~Aaij \textit{et al.} [LHCb Collaboration],
    ``Evidence for a new structure in the $J/\psi p$ and $J/\psi \bar{p}$ systems in $B_s^0 \to J/\psi p \bar{p}$ decays,''
    Phys.\ Rev.\ Lett.\ \textbf{128}, 062001 (2022).
\bibitem{Wang:2015jsa}
  Q.~Wang, X.~H.~Liu, and Q.~Zhao,
  ``Photoproduction of hidden charm pentaquark states $P_c^+(4380)$ and $P_c^+(4450)$,''
  Phys.\ Rev.\ D\ \textbf{92}, 034022 (2015).
\bibitem{Kubarovsky:2015aaa}
  V.~Kubarovsky and M.~B.~Voloshin,
  ``Formation of hidden-charm pentaquarks in photon-nucleon collisions,''
  Phys.\ Rev.\ D\ \textbf{92}, 031502 (2015).
\bibitem{Karliner:2015voa}
  M.~Karliner and J.~L.~Rosner,
  ``Photoproduction of exotic baryon resonances,''
  Phys.\ Lett.\ B\ \textbf{752}, 329 (2016).
\bibitem{Blin:2016dlf}
  A.~N.~Hiller Blin, C.~Fern\'andez-Ram\'\i{}rez, A.~Jackura, V.~Mathieu, V.~I.~Mokeev, A.~Pilloni, and A.~P.~Szczepaniak,
  ``Studying the P$_c$(4450) resonance in J/$\psi$ photoproduction off protons,''
  Phys.\ Rev.\ D\ \textbf{94}, 034002 (2016).
\bibitem{Eides:2019tgv}
  M.~I.~Eides, V.~Y.~Petrov, and M.~V.~Polyakov,
  ``New LHCb pentaquarks as hadrocharmonium states,''
  Mod.\ Phys.\ Lett.\ A\ \textbf{35}, 2050151 (2020).
\bibitem{Semenova:2019gzf}
  A.~N.~Semenova, V.~V.~Anisovich, and A.~V.~Sarantsev,
  ``New narrow LHCb pentaquarks as lowest antiquark-diquark-diquark systems,''
  Eur.\ Phys.\ J.\ A\ \textbf{56}, 142 (2020).
\bibitem{Burns:2021jlu}
    T.~J.~Burns and E.~S.~Swanson,
    ``Experimental constraints on the properties of $P_c$ states,''
    Eur.\ Phys.\ J.\ A\ \textbf{58}, 68 (2022).
\bibitem{Peng:2019wys}
    F.~Z.~Peng, M.~Z.~Liu, Y.~W.~Pan, M.~S\'anchez S\'anchez, and M.~Pavon Valderrama,
    ``Five-flavor pentaquarks and other light- and heavy-flavor symmetry partners of the LHCb hidden-charm pentaquarks,''
    Nucl.\ Phys.\ B\ \textbf{983}, 115936 (2022).
\bibitem{Zhang:2023czx}
    Z.~Zhang, J.~Liu, J.~Hu, Q.~Wang, and U.~G.~Mei\ss{}ner,
    ``Revealing the nature of hidden charm pentaquarks with machine learning,''
    [arXiv:2301.05364 [hep-ph]].
\bibitem{Gell-Mann:1961jim}
    M.~Gell-Mann and F.~Zachariasen,
    ``Form-factors and vector mesons,''
    Phys.\ Rev.\ \textbf{124}, 953 (1961).
\bibitem{Kroll:1967it}
    N.~M.~Kroll, T.~D.~Lee, and B.~Zumino,
    ``Neutral vector mesons and the hadronic electromagnetic current,''
    Phys.\ Rev.\ \textbf{157}, 1376 (1967).
\bibitem{Sakurai:1969}
    J.~J.~Sakurai,
    ``Currents and mesons,''
    (The University of Chicago Press, 1969).
\bibitem{Ali:2019lzf}
  A.~Ali \textit{et al.} [GlueX Collaboration],
  ``First measurement of near-threshold $J/\psi$ exclusive photoproduction off the proton,''
  Phys.\ Rev.\ Lett.\ \textbf{123}, 072001 (2019).
\bibitem{GlueX:2023}
  S.~Adhikari \textit{et al.} [GlueX Collaboration],
  ``Measurement of the $J/\psi$ photoproduction cross section over the full near-threshold kinematic region,''
  arXiv:2304.03845 [nucl-ex],
\bibitem{Azimov:2009ta}
  Y.~Azimov,
  ``Quantum interference of particles and resonances,''
  J.\ Phys.\ G\ \textbf{37}, 023001 (2010).
\bibitem{LeviSetti:1973}
   R.~Levi~Setti and T.~Lasinski, ``Strongly Interacting Particles (Chicago lectures in physics),''
    (The University of Chicago Press, 1973).
\bibitem{Landau:1991wop}
  L.~D.~Landau and E.~M.~Lifshitz,
  ``Quantum Mechanics : Non-Relativistic Theory,''
  (Butterworth-Heinemann, 1977).
\bibitem{Strakovsky:1984hb}
  I.~I.~Strakovsky, A.~V.~Kravtsov, and M.~G.~Ryskin,
  ``The study of the $\pi^+d\to pp$ process within the framework of $J\le4$ phase shift analysis,''
  Yad.\ Fiz.\ \textbf{40}, 429 (1984) [Sov.\ J.\ Nucl.\ Phys.\ \textbf{40}, 273 (1984)].
\bibitem{Du:2020bqj}
  M.~L.~Du, V.~Baru, F.~K.~Guo, C.~Hanhart, U.~G.~Mei\ss{}ner, A.~Nefediev, and I.~Strakovsky,
  ``Deciphering the mechanism of near-threshold $J/\psi$ photoproduction,''
  Eur.\ Phys.\ J.\ C\ \textbf{80}, 1053 (2020).
\bibitem{Strakovsky:2019bev}
  I.~Strakovsky, D.~Epifanov, and L.~Pentchev,
  ``J/$\psi$p scattering length from GlueX threshold measurements,''
  Phys.\ Rev.\ C\ \textbf{101}, 042201 (2020).
\bibitem{Pentchev:2020kao}
  L.~Pentchev and I.~I.~Strakovsky,
  ``$J/\psi$-$p$ scattering length from the total and differential photoproduction cross sections,''
  Eur.\ Phys.\ J.\ A\ \textbf{57}, 56 (2021).
\bibitem{Feinberg:1980yu}
  E.~L.~Feinberg,
  ``Hadron clusters and half dressed particles in Quantum Field Theory,''
  Usp.\ Fiz.\ Nauk\ \textbf{132}, 225 (1980) [Sov.\ Phys.\ Usp.\ \textbf{23}, 629 (1980)].
\bibitem{Strakovsky:2021vyk}
    I.~I.~Strakovsky, W.~J.~Briscoe, L.~Pentchev, and A.~Schmidt,
    ``Threshold Upsilon-meson photoproduction at the EIC and EicC,''
    Phys.\ Rev.\ D\ \textbf{104}, 074028 (2021).
\end{thebibliography}
\end{document}